\documentclass[11pt]{amsart}
\usepackage{latexsym,amssymb,calc,amsmath,amsthm,amssymb,color,graphicx,graphics}
\usepackage{bbold, epsfig, euscript,enumitem}

\usepackage[windows]{umlaute}

\usepackage[T1]{fontenc}
\usepackage{fancyhdr}

\newcommand{\esssup}[1]{\mathop{\rm ess\ sup}}
\newcommand{\essinf}[1]{\mathop{\rm ess\ inf}}
\newcommand{\N}{{\rm I\kern - 2.5pt N}}
\newcommand{\Z}{{\rm Z\kern - 5.5pt Z}}
\newcommand{\Q}{{\rm I\kern - 5.25pt Q}}
\newcommand{\C}{{\rm I\kern - 6.25pt C}}
\newcommand{\R}{{\rm I\kern - 2.5pt R}}

\newcommand{\gbf}{\mathbf{g}}
\newcommand{\Ibf}{\mathbf{I}}

\newcommand{\Jbf}{\mathbf{J}}
\newcommand{\nbf}{\mathbf{n}}

\newcommand{\Sbf}{\mathbf{S}}

\newcommand{\vbf}{\mathbf{v}}
\newcommand{\xbf}{\mathbf{x}}

\newcommand{\na}{\nabla}
\newcommand{\pa}{\partial}
\begin{document}
\title[Influence of non-ionic surfactant on mass transfer]{A continuum thermodynamic model of the influence of non-ionic surfactant on mass transfer from gas bubbles}
\author[Dieter Bothe]{Dieter Bothe \vspace{0.1in}}
\address{Department of Mathematics and Profile Area Thermofluids \&  Interfaces\\
Technical University of Darmstadt\\
Alarich-Weiss-Str.~10\\
D-64287 Darmstadt, Germany}
\email{bothe@mma.tu-darmstadt.de}

\author[Akio Tomiyama]{Akio Tomiyama \vspace{0.1in}}
\address{Department of Mechanical Engineering, Graduate
School of Engineering\\
Kobe University\\
1-1 Rokkodai, Nada, Kobe, Hyogo 657-8501, Japan}
\email{tomiyama@mech.kobe-u.ac.jp}
\date{May 1, 2025}
\begin{abstract}
\noindent
Mass transfer of gaseous components from rising bubbles to the ambient liquid depends not only on the chemical potential difference of the transfer component but also on the interfacial free energy and composition. The latter is strongly affected by surface active agents that are present in many applications. Surfactants lead to local changes in the interfacial tension, which influence the mass transfer rates in two different ways. On the one hand, inhomogeneous interfacial tension leads to Marangoni stress, which can strongly change the local hydrodynamics. One the other hand, the coverage by surfactant molecules results in a mass transfer resistance. This hindrance effect is not included in current continuum physical models.
The present work provides the experimental validation of a recently introduced extended sharp-interface model for two-phase flows with mass transfer that also accounts
for the mass transfer hindrance due to adsorbed surfactant. The crucial feature is to account for area-specific concentrations not only of adsorbed constituents but also of transfer species, and to model mass transfer as a series of two bi-directional sorption-type bulk-interface exchange processes. The resulting model is shown to quantitatively describe experimental measurements on mass transfer reduction for the dissolution of CO$_2$ bubbles in different surfactant solutions.
\end{abstract}
\maketitle
\vspace{0.1in}\\
\section{Introduction}
\noindent
Multicomponent two-phase fluid systems are ubiquitous in nature, science and engineering.
Prominent examples are droplets in the atmosphere, forming clouds or fog,  the spray of water droplets in breaking waves, gas bubbles rising through water in rivers, natural lakes or oceans, or bubbles being dispersed in other liquids within technical contact apparatuses such as bubble columns or extraction columns.
In all these examples, two bulk fluid phases are in contact at a deformable interface which is free to move.
The presence of such an interface results from either no or partial miscibility on the molecular scale of two different fluids in contact, e.g.\ oil and water, or the coexistence of a liquid and its own vapour.
In the vast majority of such two-phase fluid systems, at least two different chemical constituents (species) are present; the only exception would be a liquid/vapour system without any impurities.
In fact, the prototypical case is that of several chemical species being mixed on the molecular scale. For instance, in case of an air bubble in water, the list of involved constituents includes water, nitrogen, oxygen and carbon dioxide,
among others. Consequently, the generic two-phase fluid system is composed of two multicomponent mixtures, which form the two bulk phases, being in contact at their common interface.

Two-phase fluid systems which are out of equilibrium exchange mass, momentum and energy.
In multicomponent systems away from chemical equilibrium, matter will be exchanged across the interface,
i.e.\ a transfer of chemical constituents takes place.
Any such process in which a certain species is exchanged is termed 'mass transfer' in Chemical Engineering, a notion which we also follow here. Mass transfer occurs in all of the examples given above, such as ocean-atmospheric exchange of gaseous components (including CO$_2$ as a most relevant topic)
\cite{2b, Veron, Hall}, cloud physicochemistry \cite{tilgner2021},
gas scrubbing processes \cite{charpentier1981},
aeration for oxygen supply in bio-reactors, e.g.\ for waste water treatment \cite{Tramper, Rosso},
reactive bubble column processes \cite{Deckwer, 1b, Schlüter2021} etc.

In particular in two-phase chemical reactors, the transfer of chemical species is the necessary prerequisite
in order for the desired chemical reactions to occur. In fact, this mass transfer is often the limiting step of the overall process. This is even more true for extraction processes, in which mass transfer is the core process step to be performed.
Therefore, detailed and fundamental quantitative knowledge on the local mass transfer across fluid interfaces is of utmost importance for process control and intensification.
Besides external operating conditions which can be used to influence, say, the size and shape of bubbles or droplets, the flow conditions, pressure and temperature, small-scale interfacial phenomena can have a significant impact on the mass transfer rates. In particular,
it is observed in many applications and experiments that surface active substances (so-called surfactants)
are present, as impurities in contaminated systems or as additives, brought intentionally into the fluid system in order
to change certain properties (see, e.g., \cite{Rosen, Chowdhury, Singh, Palmer}) or being an educt or product in a chemical reaction network. Let us note in passing that even certain fluorescence tracers used for local concentration measurements turn out to be surface active, thus influencing the process that is to be monitored \cite{Weiner2019}.
Surfactants adsorb to the fluid interface, changing its interfacial free energy and interfacial tension.
Even at very small surfactant concentrations in the range of a few ppm or less within the bulk, their interfacial
concentration will typically be large and lead to significant changes in the macroscopic system behavior; see \cite{Takagi} for a review.
A classical example is the strong impact of surface active substances on the rise velocity of bubbles in aqueous systems, slowing them down considerably.

Adsorption of surfactant strongly influences the transfer rates of, e.g., gaseous components into a liquid phase.
It is commonly accepted that there are two different ways in how this influence is mediated.
First, since surfactant is inhomogeneously distributed along the interface, the interfacial tension displays non-zero
surface gradients. This induces so-called Marangoni stresses which act as tangential forces at the interface, thus changing the flow field locally, resulting for instance in the above mentioned deceleration of a rising bubble; cf.\ \cite{pesci2018computational}.
This in turn alters the ratio between diffusive and convective transport time scales, changing the diffusive mass transfer fluxes at the interface. Experimental results on the influence of Marangoni stress on mass transfer are reported, e.g., in \cite{Alves2005, Garcia2010, Jimenez2014, Huang2017, Nekoeian2019}.
Second, even in the absence of fluid flow, the partial coverage of the interface with surfactant molecules constitutes a barrier against the passage of other molecules across the interface. This leads to an additional hindrance effect, also referred to as mass transfer resistance, sometimes also referred to as a steric effect. Experimental results on this hindrance effect can be found, e.g., in
\cite{Sardeing2006, Hebrard2009, Aoki2015, Hori2019, Hori2020}.
Let us note that, similar to mass transfer, the rate of evaporation is influenced by the presence of surfactants \cite{Langmuir, Barnes1986}.
Already in \cite{Langmuir}, Langmuir introduced an energy barrier model to explain
this phenomenon, at least qualitatively. Despite this classical work, how to derive and incorporate a local barrier/hindrance effect in a thermodynamically consistent way within a rigorous continuum physical framework has been an open question until recently.
In fact, a complete and thermodynamically consistent model has been introduced in \cite{Bo-interface-mass},
and the main aim of the present paper is to provide an experimental validation of this novel mass transfer model.

In Chemical Engineering, mass transfer is usually described by the dimensionless Sherwood number Sh or the mass transfer coefficient $k_L$.
Recall that the rate of change of the molar mass $N_i$ of a species $A_i$ across an interface $\Sigma$ due to
mass transfer is given as
\begin{equation}\label{eq:mass-transfer1}
\frac{d N_i}{d t} = \int_\Sigma D_i \nabla c_i \cdot {\bf n}\, do,
\end{equation}
where $D_i$ denotes the diffusivity of species $A_i$, having molar concentration $c_i$, and the mass transfer is assumed to be purely diffusive (cf.\ below). Throughout, ${\bf n}$ denotes the outer unit normal to the interface. If $\Sigma$ has the area $A$, equation \eqref{eq:mass-transfer1}
can be recast as
\begin{equation}\label{eq:mass-transfer2}
\frac{d N_i}{d t} = A\, D_i \, \langle \frac{\partial c_i}{\partial {\bf n}} \rangle_\Sigma,
\end{equation}
where the last term denotes the interfacial average of the normal derivative of $c_i$.
Now, as this average gradient (more precisely: directional derivative) is unknown, it is replaced with a known, macroscopic difference quotient.
This requires introducing a correcting factor, the so-called Sherwood number:
\begin{equation}\label{eq:mass-transfer3}
\frac{d N_i}{d t} = - \, {\rm Sh}\, A\, D_i \, \frac{c_i^b - c_i^S}{L},
\end{equation}
where $c_i^b$ denotes a characteristic bulk concentration, $c_i^S$ an interface concentration and $L$ a characteristic distance from the interface at which the bulk concentration equals (approximately) $c_i^b$.
Division of \eqref{eq:mass-transfer3} by the volume $V$, which holds $N_i$, leads to the local version
\begin{equation}\label{eq:mass-transfer4}
\frac{d c_i}{d t} = - \, k_L\, a\, \big( c_i^b - c_i^S \big),
\end{equation}
in which
\begin{equation}\label{eq:Sh}
k_L=\frac{{\rm Sh} \, D_i}{L}
\end{equation}
is the so-called mass transfer coefficient and  $a:=A/V$ the specific interfacial area.

Even for single rising bubble, the mass transfer coefficient $k_L$ depends on various factors such as the diameter, shape and rise velocity of the bubble, the physical properties of the two phases, the presence of walls and, most relevant for the present paper, the presence of surface-active agents (surfactants). A number of studies have been carried out to investigate the effects of these factors on $k_L$ and corresponding $k_L$-correlations have been proposed \cite{3b}.
The mass transfer coefficient for fully-contaminated bubbles under wall effects was studied by Aoki et al.\ \cite{5b, Aoki2015, 7b}. 
Aoki et al.\ proposed $k_L$-correlations applicable to ellipsoidal and Taylor bubbles fully-contaminated with Triton X-100 \cite{Aoki2015} and with alcohols of various carbon-chain lengths \cite{7b} by taking into account the above-mentioned surfactant effects on $k_L$.

The present validation of the novel mass transfer model relies on new $k_L$ data for single CO$_2$ bubbles rising through surfactant solutions in vertical pipe of diameters, $d_P=12.5$ mm. The ratio of the sphere-volume-equivalent bubble diameter $d_B$ to the pipe diameter $d_P$ was widely varied to cover various bubble shapes: ellipsoidal, semi-Taylor and Taylor bubbles have been investigated. 
Let us note that the mass transfer model applies to the transfer of an arbitrary species across a fluid interface.
However, for reliable and accurate measurements of $k_L$ from bubble dissolution experiments, the high solubility of CO$_2$ in aqueous solutions is the reason for the choice of this material pairing.

The paper is organised as follows. For comparison, in Section~\ref{sec1} we briefly recall the current 'standard' sharp-interface continuum mechanical model \cite{danckwerts1970gas} for (species) mass transfer, being in common use. Section~\ref{sec2} introduces the new mass transfer model and briefly explains its derivation.
Section~\ref{sec3} explains the experimental setup used to evaluate mass transfer rates for CO$_2$ gas bubbles, rising in surfactant solution, and the procedure used to evaluate the measurements to obtain the $k_L$ coefficient. The obtained data is used to validate the novel mass transfer reduction model. This requires the derivation of a bubble-integral mass transfer model, the generation of 'synthetic', interpolated data and the fitting of the model parameters to see how accurate the data can be obtained from an overall model. This is the content of Section~\ref{sec4}. The paper ends with conclusions and a brief outlook.
\section{Mass transfer across fluid interfaces - the standard model}\label{sec1}
\noindent
In continuum mechanical descriptions of mass transfer across fluid interfaces, one common approach employs
the sharp-interface assumption, i.e.\ the interface between the contacting bulk phases is a surface of zero thickness.
Assuming constant density or small Mach number flows, the standard sharp-interface model is then
based on the incompressible two-phase Navier-Stokes equations for fluid systems without phase change. Inside the fluid phases, the governing equations are
\begin{align}
	&\na \cdot \vbf=0,\label{E1}\\
	&\pa_t (\rho\vbf)+ \na \cdot (\rho \vbf \otimes \vbf)+\na p=\na \cdot \Sbf^{\rm visc}+\rho\gbf, \label{E2}
\end{align}
where $\partial_t$ is short for $\frac{\partial}{\partial t}$.
We use standard notation, where $\rho$ denotes the mass density, $\vbf$ the barycentric velocity, $p$ the pressure, $\gbf$ the body force density due to gravity and $\Sbf^{\rm visc}$ the viscous stress tensor for a Newtonian fluid, i.e.\
\begin{equation}\label{E3}
\Sbf^{\rm visc}=\eta(\na\vbf+\na \vbf^{\sf T})
\end{equation}
with the dynamic viscosity $\eta >0$. Note that all quantities, in particular all material parameters, depend on the respective phase.
Whenever distinction between the different phases is necessary, $+$ and $-$ are used as phase indices.
The standard interfacial jump conditions for total mass and momentum are
\begin{equation}\label{E4}
[\![\vbf]\!]=0, \quad [\![-\Sbf]\!] \cdot \nbf_\Sigma=\sigma \kappa_\Sigma \nbf_\Sigma + \nabla_\Sigma \sigma,
\end{equation}
where $\Sbf=-p \Ibf+\Sbf^{\rm visc}$ is the stress tensor, $\sigma$ the surface tension, $\kappa_\Sigma=-\na \cdot \nbf_\Sigma$ is twice the mean curvature of the interface $\Sigma$ and $\nabla_\Sigma$ denotes the surface gradient. Moreover, $\nbf_\Sigma$ is the unit normal at the interface directed into one of the bulk phases and
\begin{equation}\label{E5}
[\![\phi]\!](\xbf)=\lim\limits_{h \to 0+} \big(\phi(\xbf+h\nbf_\Sigma)-\phi(\xbf-h\nbf_\Sigma)\big)
\end{equation}
denotes the jump of a field $\phi$ across the interface at the location $\xbf$.

The local molar concentration $c_i$ of a chemical species $A_i$ is governed by the balance equation
\begin{equation}\label{E6}
\pa_t c_i+\na \cdot (c_i \vbf+\Jbf_i)=r_i,
\end{equation}
where the molecular fluxes $\Jbf_i$ are typically modeled according to Fick's law as
\begin{equation}\label{E7}
\Jbf_i=-D_i \na c_i
\end{equation}
with constant diffusivity $D_i$. The source term on the right-hand side in equation (\ref{E6}) accounts for chemical reactions. At the interface, the diffusive fluxes in normal direction are usually supposed to be continuous, i.e.\ \begin{equation}\label{E8}
[\![\Jbf_i]\!] \cdot \nbf_\Sigma=0.
\end{equation}
One more constitutive equation is needed to determine the concentration profiles of $A_i$, where instantaneous local chemical equilibrium at the interface is usually employed. This means continuity of the chemical potentials $\mu_i$ at the interface, i.e.\ \begin{equation}\label{E9}
[\![\mu_i]\!] =0.
\end{equation}
Using standard relations for $\mu_i$ leads to Henry's law. In the form for molar concentrations, the latter states that
\begin{equation}\label{E10}
c^-_k=c^+_k/H_k
\end{equation}
with a Henry coefficient $H_k$, which is often assumed to be constant.

Throughout this paper, the superscripts $\pm$ are used to distinguish between the two bulk phases.
Equations (\ref{E1})--(\ref{E4}) and (\ref{E6})--(\ref{E10}) comprise what we call the 'standard model'. This standard model, sometimes with further simplifications like constant surface tension, homogeneous gas phase concentrations or constant liquid-sided concentration at the interface, is the basis of almost all detailed numerical simulations of mass transfer across fluid interfaces up to now; see \cite{bothe2004direct, haroun2010volume, marschall2012numerical, boulhasanzadeh2012multiscale, 1, WeBo17, maes2018new, Weiner2019, claassen2020improved, weiner2022assessment} and the extensive list of references given there.

\section{Mass transfer influenced by adsorbed surfactant - A novel continuum thermodynamical model}\label{sec2}
We briefly recap the novel mass transfer model introduced in \cite{Bo-interface-mass} together with the main ideas and steps employed in its derivation.
The model is derived within the sharp-interface framework, i.e.\ a hypersurface $\Sigma$ of zero thickness represents the thin physical layer in which the partial mass densities change from one to the other (local) bulk value.
We consider two-phase fluid systems far away from the critical point, in which case the thickness of this transition layer is in the order of a few \AA ngström. 
For instance, the water-air interface has a thickness of about 0.6-0.8 nm under normal conditions, as has been measured very recently \cite{fellows2024thick}.
Hence, at least on meso- and macroscopic scales, the sharp-interface model usually provides an excellent approximation of the physical system by replacing the continuous but extremely sharp transition in density profiles and other intensive
properties with discontinuous fields, having one-sided limits at the surface of discontinuity.

To capture mass transfer hindrance due to surface coverage, the crucial step is to allow for a possibly non-zero interfacial concentration in particular for every \emph{transfer component}, i.e.\ any $A_i$ with non-zero concentrations on both sides of the interface.
Note that any constituent $A_i$ of the mixture which is able to pass from one of the bulk phases to the opposite side of the interface
necessarily has to cross the thin transition zone between the bulk phases which is represented here by the sharp interface $\Sigma$.
Consequently, constituent $A_i$ will typically show a non-zero interfacial concentration $c^\Sigma_i >0$, hence its modeling requires a full mass balance including bulk and interface contributions as required for soluble surfactant.
Note that $c^\Sigma_i$ is an area-specific concentration and even if $c^\Sigma_i >0$ is very small, as is the case for transfer species, this concentration is relevant to account for the mixture thermodynamics on the interface. 
\begin{figure}
\centering
\includegraphics[width=2.4in]{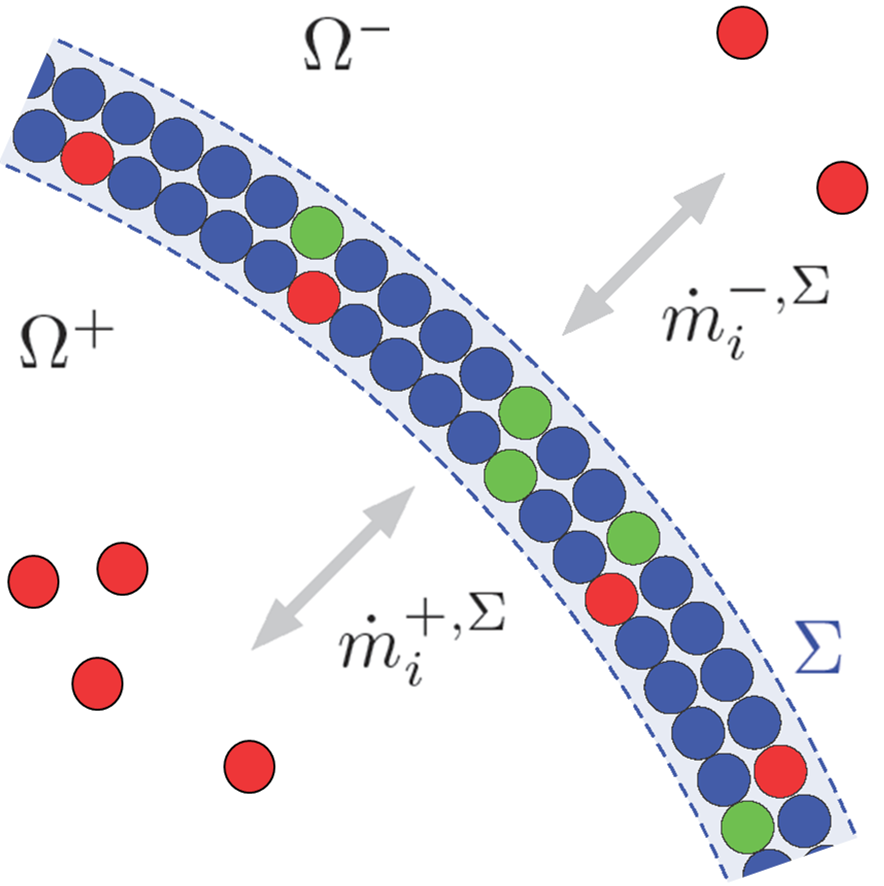}
\caption{Mass transfer as a series of two bi-directional, sorption-like sub-processes. Figure adapted from \cite{Bo-interface-mass}.}\label{Fig:one-sided-sorptions}
\end{figure}

The following is the core idea of our approach: Since the transfer of mass requires molecules to enter or leave the
interface, we consider mass transfer as the sequence of two one-side bulk-surface exchange processes, very similar to ad- and desorption processes.
This is illustrated in Figure~\ref{Fig:one-sided-sorptions} and turns out to be the key point for enabling a coupling between the local presence of a surface active species, say $A_k$, and the transfer of a non-surface active constituent, say $A_i$. Indeed, decomposing the mass transfer process
in a series of two sorption-like sub-processes changes the driving force for mass transfer (in the isothermal case, say)
from the difference $\mu_i^+ -\mu_i^-$
of one-sided limits of bulk chemical potentials to two different driving forces of the form $\mu_i^\pm -\mu_i^\Sigma$, thus introducing
the interfacial chemical potentials. The latter depend on the surface tension, which itself is strongly affected by the presence
of $A_k$, thus mediating the influence of $A_k$ via $\mu_k^\Sigma$ on the mass transfer rate of $A_i$.

Due to the above considerations, the extended model includes interface concentrations (unit: mol/m$^2$) for all constituents. Let us note in passing that the interfacial concentration of a certain species may, however, vanish locally at a specific instant of time or even globally on $\Sigma$.
The molar mass balance for $A_i$ on the interface reads as
\begin{equation}\label{E11}
\pa^\Sigma_t c^\Sigma_i + \na_\Sigma \cdot (c^\Sigma_i \vbf^\Sigma + \Jbf^\Sigma_i)+ [\![c_i(\vbf-\vbf^\Sigma)+\Jbf_i]\!] \cdot \nbf_\Sigma=r^\Sigma_i,
\end{equation}
where $\vbf^\Sigma$ denotes the interface barycentric velocity, $\Jbf^\Sigma_i$ the interfacial diffusive flux and $r^\Sigma_i$ is the total molar rate of change of $A_i$ due to interface chemical reactions between the species. Furthermore, $\pa^\Sigma_t$ denotes the time derivative along a path which follows the interface's normal motion. Observe that even for vanishing interface concentrations and without interface chemistry, equation (\ref{E11}) does not reduce to (\ref{E8}) from the standard model, but the resulting jump condition also contains a convective part due to an interface motion relative to the bulk matter. This is crucial, for example, to model the condensation of a vapor bubble or the dissolution of a pure gas bubble \cite{13}.

Concerning the balance of momentum, we keep the incompressible two-phase Navier-Stokes equations to obtain the barycentric bulk velocities.
We also restrict to isothermal conditions as this is sufficient to model the mass transfer in the absence of heating due to, say, strongly exothermic reactions. We thus focus on the species equations including the interface balances and rewrite the jump-bracket as
\begin{equation}\label{E14}
[\![c_i(\vbf-\vbf^\Sigma)+\Jbf_i]\!]  \cdot \nbf_\Sigma=- \,(\, \dot{m}^{+,\Sigma}_i + \dot{m}^{-,\Sigma}_i \,),\vspace{-0.05in}
\end{equation}
where
\begin{equation}\label{E15}
\dot{m}^{\pm,\Sigma}_i= (c^\pm_i(\vbf^\pm-\vbf^\Sigma)+ \Jbf^\pm_i) \cdot \nbf^\pm,
\end{equation}
with $\nbf^\pm$ denoting the outer unit normal to the respective bulk phase. 

The terms $\dot{m}^{\pm,\Sigma}_i$ represent the central object of mass transfer modeling as they denote the (molar) mass transfer rates of constituent $A_i$ from the respective bulk phase to the interface.
To obtain a realistic closure for $\dot{m}^{\pm,\Sigma}_i$, the one-sided but bi-directional mass transfer terms are split into two uni-directional (adsorption and a desorption-like) terms according to
\begin{equation}\label{E16}
\dot{m}^{+,\Sigma}_i=s^{ad,+}_i -s^{de,+}_i, \quad \dot{m}^{-,\Sigma}_i= s^{ad,-}_i -s^{de,-}_i.
\end{equation}
The transfer of (molar) mass is accompanied with the production of entropy and any thermodynamically consistent closure must guarantee positivity of the associated entropy production rate.
Neglecting a kinetic and a viscous term inside the mass transfer entropy production rate, see \cite{Bo-interface-mass} for details, the contributions
\begin{equation}\label{E48}
\zeta^\pm_{\rm TRANS}= \sum\limits_{i=1}^N \big(s^{ad,\pm}_i- s^{de,\pm}_i\big) \big(\frac{\mu^\pm_i}{T^\pm} - \frac{\mu^\Sigma_i}{T^\Sigma}\big)
\end{equation}
need to be non-negative. 

We now add the realistic assumption of a continuous temperature field, i.e.\ $T^\pm_{|\Sigma} =T^\Sigma$; from here on, we just write $T$ for the temperature field throughout the bulk phase and on the interface.
Ignoring direct mass transfer cross-effects between different species, we thus need to guarantee
\begin{equation}\label{E48b}
\big(s^{ad,\pm}_i- s^{de,\pm}_i\big) \big( \mu^\pm_i  -  \mu^\Sigma_i \big) \geq 0 \qquad \mbox{ for all } i.
\end{equation}
Consequently, the one-sided, uni-directional sorption-like transfer rates $s^{ad,+}_i$, $s^{de,+}_i$, $s^{ad,-}_i$ and $s^{de,-}_i$
need to be appropriately modeled via material-depen\-dent constitutive relations.
Here it is crucial to employ a nonlinear closure, very similar as for chemical reactions, since the system can be far away
from local chemical equilibrium at interface points regarding the interfacial chemical potential and the adjacent one-sided limits of the bulk chemical potentials. The appropriate closure relation reads as
\begin{equation}\label{E49}
\ln \frac{s^{ad,\pm}_i}{s^{de,\pm}_i}= \frac{\alpha^\pm_i}{RT} \big(\mu^\pm_i-\mu^\Sigma_i \big) \quad\text{ with }\quad \alpha^\pm_i \geq 0,
\end{equation}
where $\mu^\pm_i$ are the bulk chemical potentials and $\mu^\Sigma_i$ is the interfacial chemical potential.
As we formulate the transfer rates for the molar mass, the chemical potentials are also molar-based.
Note that the molar-based chemical potential has the same physical unit as $RT$, while
the mass-based chemical potentials contain the factor $M_i$, the molar mass of $A_i$.

At this point, information about the chemical potentials of the fluid system is required.
For the bulk phases, we use the general representation
\begin{equation}\label{E52}
\mu^\pm_i(T,p,x_k)= g^\pm_i(T,p)+ RT \ln a^\pm_i
\end{equation}
with a reference chemical potential $g^\pm_i(T,p)$ and the so-called activity $a^\pm_i$ of $A_i$ in the respective bulk phase.
The activity $a_i$ is often written as $\gamma_i x_i$ with $x_i=c_i/c$ the molar fraction of $A_i$ and $\gamma_i$ the activity coefficient. 
Note that \eqref{E52} is just a different way to write the general chemical potential, since the activity is allowed to depend on the full set of state variables $(T,p,x_k)$.
This representation is useful as it resembles the case of ideal mixtures, for which $g^\pm_i(T,p)$
is the Gibbs free energy of component $A_i$ under the temperature and pressure of the mixture, while $a_i=x_i$ (i.e.\ the activity coefficient satisfies $\gamma_i =1$).

The interfacial chemical potentials will be derived from an appropriate interface free energy model,
which is associated to the \emph{Langmuir adsorption isotherm} \cite{Kralchevsky} and is given by
\begin{equation}\label{surface-free-energy-Langmuir}
\big( c^\Sigma \psi^{\Sigma} \big) (T,c_1^\Sigma, \ldots, c_N^\Sigma)=
- p_0^\Sigma (T) + \sum_{k=1}^N \alpha_k^\Sigma (T) c_k^\Sigma + RT \sum_{k=0}^N c_k^\Sigma \ln \theta_k,
\end{equation}
where $c^\Sigma=\sum_{k=1}^N c_k^\Sigma$ is the total molar concentration on the interface, $\psi^{\Sigma}$ the specific (per mole) interfacial free energy, and $\theta_k:=c_k^\Sigma / c_\infty^\Sigma$ with some $c_\infty^\Sigma >0$ that characterises the capacity of the interface to locally host adsorbed or transferring molecules. In \eqref{surface-free-energy-Langmuir}, the additional constituent $c_0^\Sigma$ stands for the remaining (local) capacity
of the interface to carry further molecules, corresponding to free sites in case of a lattice.
To stress the relevance of the term $c_0^\Sigma \ln \theta_0$, we rewrite \eqref{surface-free-energy-Langmuir} as
\begin{equation}\label{surface-free-energy-Langmuir2}
c^\Sigma \psi^{\Sigma} =
- p_0^\Sigma (T) + \sum_{k=1}^N \alpha_k^\Sigma (T) c_k^\Sigma + RT \sum_{k=1}^N c_k^\Sigma \ln \theta_k
+ c_\infty^\Sigma RT (1-\theta) \ln (1-\theta)
\end{equation}
with $\theta = \sum_{k=1}^N \theta_k=c^\Sigma / c^\Sigma_\infty$ denoting the total (local) coverage of $\Sigma$.
Differentiation of $c^\Sigma \psi^{\Sigma} =  c^\Sigma \psi^{\Sigma} (T, c_1^\Sigma, \ldots,  c_N^\Sigma)$
from  \eqref{surface-free-energy-Langmuir2} w.r.\ to $c_k^\Sigma$
yields
\begin{equation}\label{surface-chem-pot-Langmuir}
\mu_k^{\Sigma} =  \alpha_k^\Sigma (T) + RT \ln \theta_k
- RT  \ln (1- \theta ).
\end{equation}
The interfacial Euler relation in the local form, i.e.\
\begin{equation}
c^\Sigma \psi^{\Sigma} + p^\Sigma = \sum_{k=1}^N c_k^\Sigma \mu_k^{\Sigma},
\end{equation}
yields the surface pressure as
\begin{equation}\label{surface-pressure-Langmuir}
p^\Sigma = p_0^\Sigma (T) - c_\infty^\Sigma RT  \ln (1- \theta ).
\end{equation}
Note that $\sigma:= - p^\Sigma$ is the interfacial tension and that $p^\Sigma \to \infty$ as $\theta \to 1-$ as expected.
Equation \eqref{surface-pressure-Langmuir} allows to rewrite the surface chemical potential as
\begin{equation}\label{surface-chem-pot-Langmuir2}
\mu_k^{\Sigma} =  \alpha_k^\Sigma + \frac{p^\Sigma-p_0^\Sigma (T)}{c_\infty^\Sigma} + RT \ln \theta_k.
\end{equation}
This is a form, similar to that of an ideal mixture, since
\begin{equation}\label{surface-chem-pot-Langmuir3}
\mu_k^{\Sigma} =  g_k^\Sigma (T, p^\Sigma) + RT \ln \theta_k
\end{equation}
with
\begin{equation}\label{E62b}
g_k^\Sigma (T, p^\Sigma):= \alpha_k^\Sigma + (p^\Sigma-p_0^\Sigma (T))/c_\infty^\Sigma.
\end{equation}
Indeed, \eqref{surface-chem-pot-Langmuir3} resembles the form of chemical potentials in ideal mixtures
except for the use of $\theta_k$ instead of the molar fraction.
But note that in the limit as $c_k^\Sigma \to c_\infty^\Sigma -$, both
$g_k^\Sigma (T, p^\Sigma)$ and $\mu_k^{\Sigma}$ become infinite, since the surface pressure diverges.
Hence we cannot conclude that $g_k^\Sigma (T, p^\Sigma)$ is the surface Gibbs free energy of the pure substance $A_k$
(in its adsorbed form).

One of the rates, either the ad- or the desorption rate, still has to be modeled based on a micro-theory or experimental knowledge, while the other rate then follows from (\ref{E49}).
Desorption is often more easy to model, where the simplest rate function is
\begin{equation}\label{E50}
s^{de,\pm}_i= k^{de,\pm}_i \theta_i.
\end{equation}
Insertion of the chemical potentials from \eqref{E52} and \eqref{surface-chem-pot-Langmuir3} into (\ref{E49}) then yields
\begin{equation}\label{E54}
s^{ad,\pm}_i= k^{de,\pm}_i \exp \big(\frac{g^\pm_i-g^\Sigma_i}{RT} \big) a^\pm_i,
\end{equation}
where the simplest choice of $\alpha^\pm_i=1$ has been used; the latter will turn out to already be sufficient to describe mass transfer hindrance in a realistic manner.

Next, it is assumed that (\ref{E11}) can be approximated by
\begin{equation}\label{E55}
[\![\dot{m}_i]\!]=0 \quad \Leftrightarrow \quad \dot{m}^{+,\Sigma}_i+ \dot{m}^{-,\Sigma}_i=0.
\end{equation}
While this will automatically hold if the interface concentration $c_i^\Sigma$ of $A_i$ is sufficiently small, the latter is not a necessary prerequisite.
For the equation \eqref{E55} to hold at a good accuracy, it is required that the flux of $A_i$ inside the interface is small compared to the flux of $A_i$ from the bulk into the interface or vice versa.
Then, employing (\ref{E16}), (\ref{E50}) and (\ref{E54}), it follows that
\begin{equation}\label{E57}
k^{de,+}_i \left(\exp \big( \frac{g^+_i-g^\Sigma_i}{RT}\big)a^+_i-\theta_i\right)+ k^{de,-}_i \left(\exp \big( \frac{g^-_i-g^\Sigma_i}{RT}\big)a^-_i-\theta_i\right)=0,
\end{equation}
hence
\begin{equation}\label{E58}
\displaystyle
\theta_i= \frac{k^{de,+}_i \exp \big(\frac{g^+_i-g^\Sigma_i}{RT}\big)a^+_i
+ k^{de,-}_i \exp \big( \frac{g^-_i-g^\Sigma_i}{RT}\big)a^-_i}{k^{de,+}_i+ k^{de,-}_i}.
\end{equation}
Inserting this expression into the first term in (\ref{E57}), equation (\ref{E16}) yields
\begin{equation}\label{E59}
\dot{m}^{+,\Sigma}_i(=- \dot{m}^{-,\Sigma}_i)= k_i \exp \big(- \frac{g^\Sigma_i}{RT}\big) \left( \exp \big( \frac{g^+_i}{RT}\big) a^+_i- \exp \big( \frac{g^-_i}{RT}\big)a^-_i\right)
\end{equation}
with the mass transfer rate coefficient 
\begin{equation}\label{E59-k}
k_i = \frac{k^{de,+}_ik^{de,-}_i}{k^{de,+}_i+ k^{de,-}_i}.
\end{equation}
%

For comparison, note that an analogous modeling but without interface concentrations yields
\begin{equation}\label{E61}
\dot{m}^{+,\Sigma}_i(=- \dot{m}^{-,\Sigma}_i)= k_i \exp \big( -\frac{g^-_i}{RT} \big) \left(\exp \big(\frac{g^+_i}{RT}\big) a^+_i- \exp\big(\frac{g^-_i}{RT}\big)a^-_i\right).
\end{equation}
The most important difference is that in (\ref{E59}) the mass transfer rate is influenced by the interfacial tension via the interface Gibbs free energy, i.e.\ via $g_i^\Sigma$, which accounts for the effect of surfactants on the mass transfer of the considered transfer component.

Insertion of (\ref{E62b}) into (\ref{E59}) implies the relation
\begin{equation}
\dot{m}^{+,\Sigma}_i=  k_i
\exp \big( \frac{- \, p^\Sigma}{c_i^{\Sigma,\infty} RT}\big)
\left( \exp \big( \frac{g^+_i}{RT}\big) a^+_i- \exp \big( \frac{g^-_i}{RT}\big)a^-_i\right).
\end{equation}
Taking the clean surface as the reference state, this yields
\begin{equation}\label{E65b}
\dot{m}^{\rm contam}_i= 
\exp \Big( - \frac{\sigma_{\rm clean} - \sigma_{\rm contam}}{c_i^{\Sigma,\infty} RT}\Big)
\dot{m}^{\rm clean}_i.
\end{equation}
The mass transfer reduction in (\ref{E65b}) corresponds to an exponential
damping factor of Boltzmann type, i.e.\ a factor
of the form $k \exp (-a \, p^\Sigma /RT)$, in accordance with the energy barrier model due to Langmuir; see \cite{Langmuir}, \cite{Ciani} and the references given there. In the present model, this term is incorporated into a complete and thermodynamically consistent model.

For full details of the model derivation we refer to \cite{Bo-interface-mass}, \cite{Bo-interface-mass-II}, while
for fundamentals on continuum mechanical modeling of two-phase fluid systems in general we also refer
in particular to \cite{3, 4, Bedeaux, 5}.
\section{Experimental study of mass transfer hindrance}\label{sec3}
\begin{figure}
\centering
\includegraphics[width=3.4in]{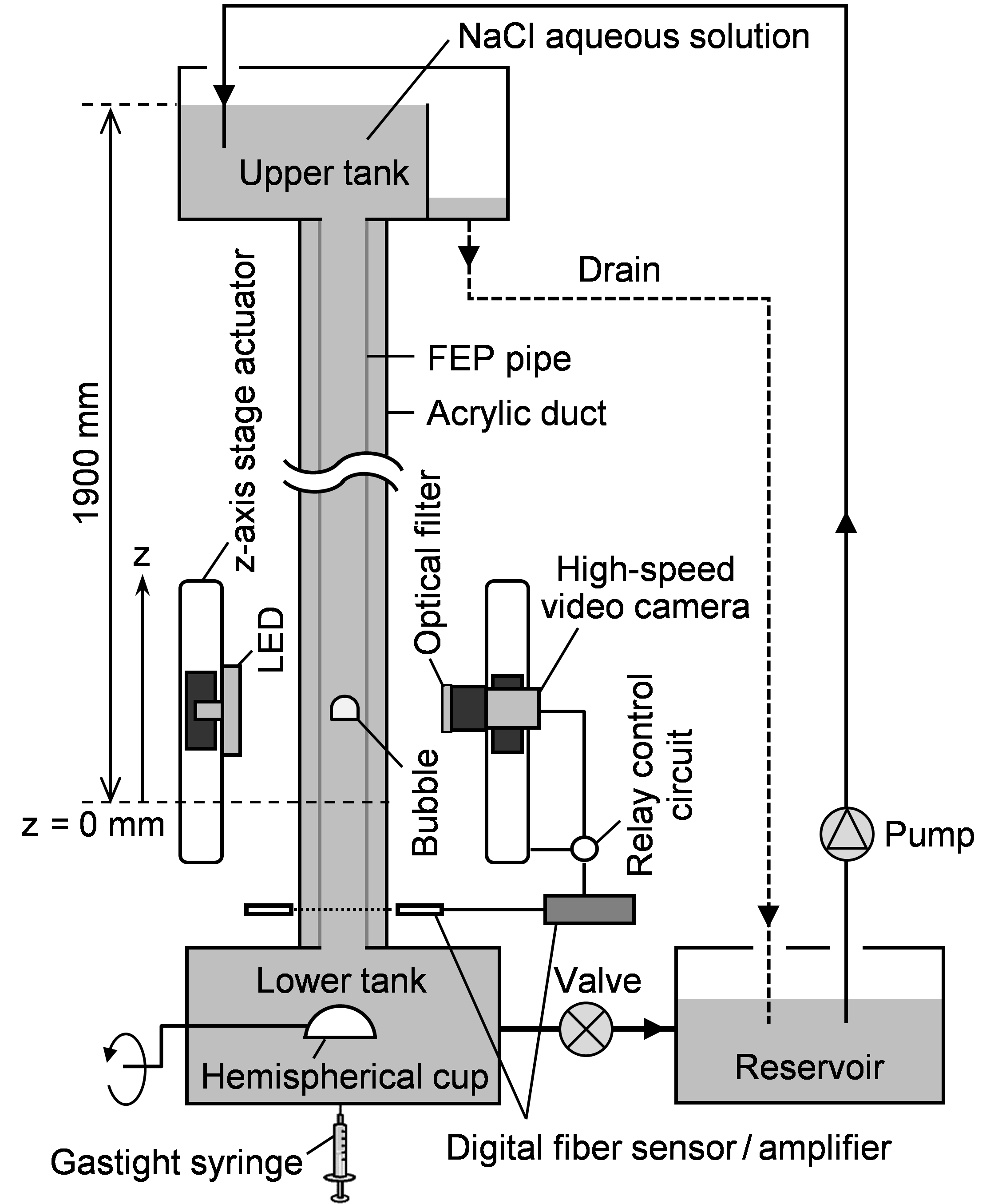}
\caption{Experimental setup for mass transfer measurements.}\label{Fig:exp-setup}
\end{figure}

\noindent
%
We study the mass transfer of CO$_2$ from rising gas bubbles into water and surfactant solutions. The mass transfer coefficient $k_L$ will be obtained from
the rate of volume change of the dissolving bubbles and in order to obtain maximum dissolution rates, pure CO$_2$ bubbles are employed.
Fig.~\ref{Fig:exp-setup} shows the experimental setup, which consists of the vertical pipe, the lower and upper tanks and the reservoir. The upper tank and the reservoir were opened to the atmosphere, and therefore, the gas components in the liquid were in their equilibrium to the atmosphere. A pipe with diameter of $d_P$ = 12.5 mm and a length of 2000 mm was used. The reference elevation ($z$ = 0 mm) was set at 1900 mm below the water surface in the upper tank. The pipe was made of fluorinated-ethylene-propylene (FEP) resin. The FEP pipe was installed in the acrylic duct. Water was filled in the gap between the duct and the pipe to reduce optical distortion in bubble images. Water purified using a Millipore system (Elix 3) and CO$_2$ of 99.9 vol.\% purity were used for the liquid and gas phases, respectively. The physical properties of the liquid and gas phases are shown in Table \ref{tab:1}, where $\rho_L$ is the liquid density, $\mu_L$ the liquid viscosity, $\sigma$ the surface tension, $D_L$ the diffusion coefficient of CO$_2$ in the liquid phase \cite{9b}, and $C^S$ the CO$_2$ concentration at the bubble surface \cite{10b}, where we omit a species index. 
\begin{table}[ht!]
\caption{\label{tab:1} Physical properties of solvent}
\begin{tabular}{lllll}
\hline 
$\rho_L$ [kg/m$^3$]	&	$\mu_L$ [Pa $\cdot$ s]	&	$\sigma$ [mN/m]		&	$D_L$ [m$^2$/s] 	&	$C^S$ [mol/m$^3$] \vspace{1pt}\\ \hline \\[-2ex]
997		&	$0.89 \times 10^{-3}$		& 72.5 	&	$1.90 \times 10^{-9}$	& 34.0	 \\ \hline
\end{tabular}
\end{table}
%
%

Experiments were carried out at atmospheric pressure and room temperature (298$\pm$1.0 K). The liquid in the pipe was refreshed before each run by circulating the liquid using a pump. A predetermined amount of CO$_2$ gas was injected from the bottom of the lower tank and stored in the hemispherical cup by using the gastight syringe. A single bubble was released by rotating the cup. Front and side images of a bubble in the test section were recorded using the two synchronized video cameras (Integrated Design Tool, M3, frame rate: 250 frame/s, exposure time: 1000 $\mu$s, spatial resolution: 0.04--0.05 mm/pixel), which were mounted on the $z$-axis actuators (SUS Corp., SA-S6AM). The green and red LED light sources (NICHIA, NSPG510AS; ROHM, SLI-580UT3F) were used for back illumination. The motion of the cameras and the LED lights were synchronized using the actuators. Bubbles were tracked for $0 \leq z \leq 550 \,\rm{mm}$.

An image processing method \cite{5b} was utilized to measure bubble volumes, diameters and positions. The original gray-scale images were transformed into binary images. By assuming that the horizontal cross sections of a bubble were elliptical, a three-dimensional bubble shape was reconstructed by piling up the elliptic disks. The bubble diameter $d_B$ was evaluated from the volume of the reconstructed bubble shape. The rise velocities $v_B$ of the bubbles in the stagnant liquids were calculated from the rates of change in the axial bubble position. Uncertainties estimated at 95\% confidence in $d_B$ and $v_B$ were $\pm$2.1\% and 0.20\%, respectively. The $d_B$ ranged from 5 to 15 mm, hence the dimensionless diameters $x_B:=d_B/d_P$ were in the range 0.4--1.2.
The Sherwood number Sh from \eqref{eq:mass-transfer3} and the mass transfer coefficient $k_L$ from \eqref{eq:Sh}
were evaluated from the rate of decrease in $d_B$. The details of the evaluation procedure can be found in \cite{Hori2020}.
\begin{figure}
\centering
\includegraphics[width=4.0in]{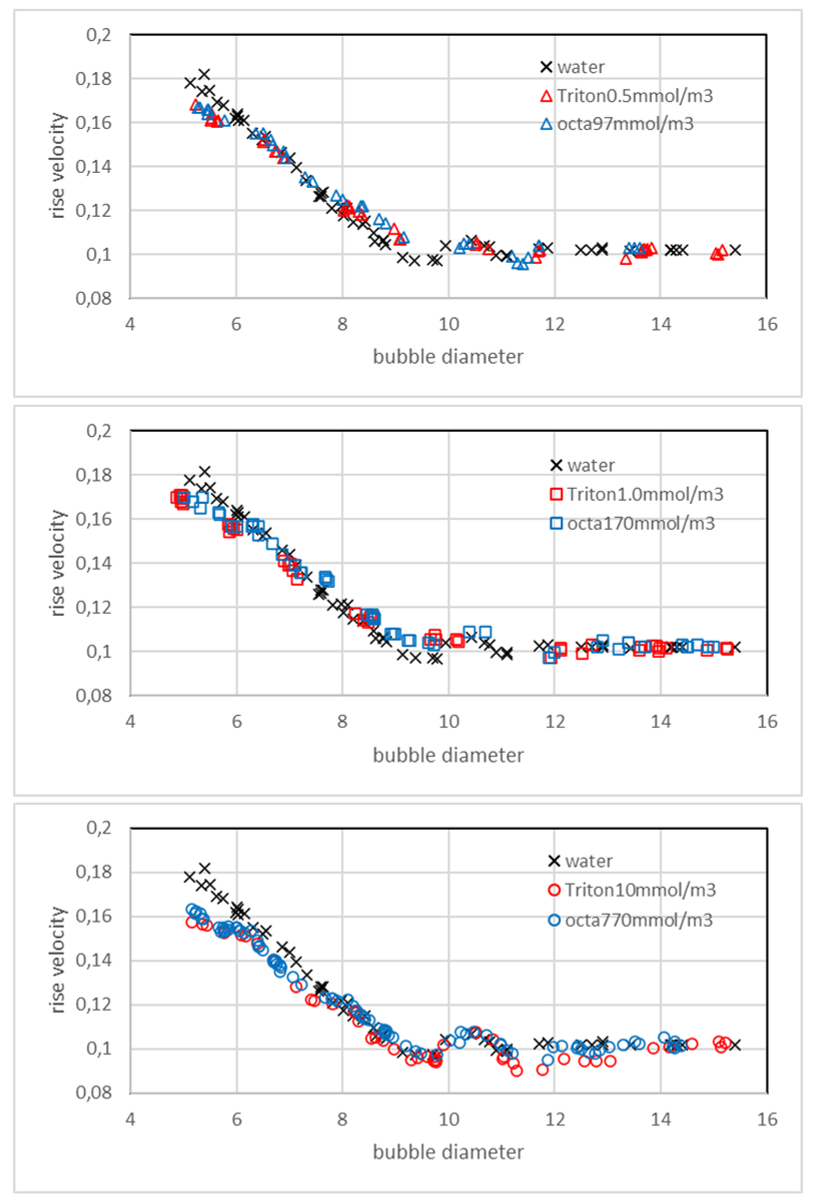}
\caption{Bubble rise velocities in m/s vs.\ bubble diameter in mm for water and different surfactant solutions. Low surfactant concentrations (top), medium concentrations (medium) and high concentrations (bottom).}\label{Fig:rise-velocities}
\end{figure}
\begin{figure}
\centering
\includegraphics[width=4.2in]{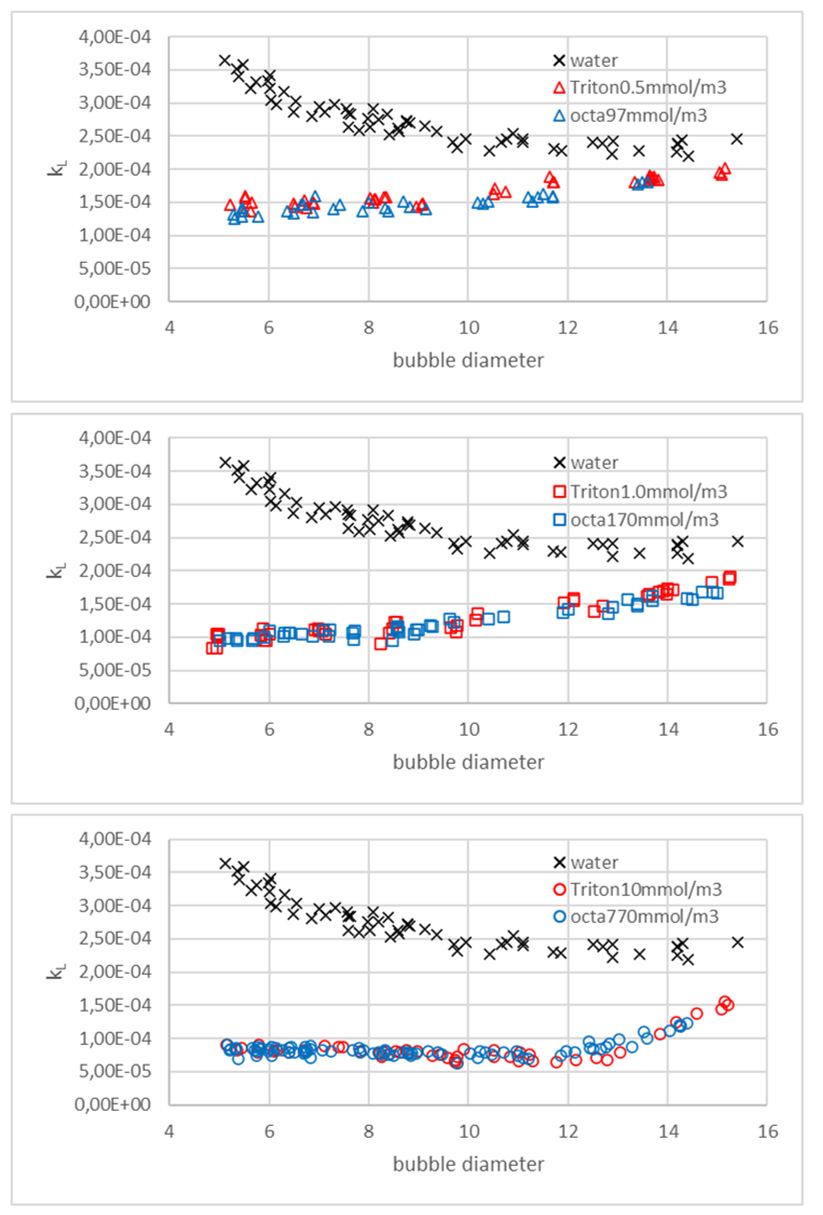}
\caption{Mass transfer coefficient $k_L$ in m/s vs.\ bubble diameter in mm for water and different surfactant solutions. Low surfactant concentrations (top), medium concentrations (medium) and high concentrations (bottom).}\label{Fig:kL-all-conc}
\end{figure}

Figure~\ref{Fig:rise-velocities} displays the rise velocities of the different bubbles, which depend on the bubble diameter and on the surfactant concentration. Each panel in this figure shows the rise velocities for bubbles in clean water together with bubble rise velocities for surfactant solutions of two different surfactants (1-octanol and Triton X-100). The surfactant concentrations are grouped in 'low', 'medium' and 'high', where the physical properties are listed in Tab.~\ref{tab:2}. The individual surfactant concentrations within each group are adjusted, so as to yield the same surface tension. 
\begin{table}[ht!]
\caption{\label{tab:2} Physical properties of surfactant solutions}
\begin{tabular}{lllll}
\hline
group					&	clean					&	low \quad						&	medium 								&	high \\ \hline
$c_{\rm oct}$ [mmol/m$^3$] 						&	0							&	97 									&	170 									&	770 \\ \hline
$c_{\rm Tri}$ [mmol/m$^3$] 						&	0							&	0.5 									&	1.0 									&	10 \\ \hline
$\sigma \;\;\;$ [mN/m]					& 72 								&	68 								&	65 								&	52\\ \hline
\end{tabular}
\end{table}
The dependence on the diameter shows a 'linear' (more precisely, an affine) relation for bubble diameters $d_B$ up to about 9 mm, corresponding to a dimensionless diameter up to $x_B =0.72=9/12.5$. For larger bubble diameters, the rise velocity is almost constant, except for a moderate variation for the surfactant solutions of highest concentration. The different relations are accompanied with different bubble shapes.
For bubbles of $x_B\leq 0.72$, the bubbles show ellipsoidal shapes and rise velocities increase as the diameter increases, mainly due to the wall effect. For $x_B$ somewhat larger than 0.72, the bubble shapes are influenced by the confinement due to the pipe walls and form the typical Taylor bubble shapes.

For each rising bubble, evaluation of the monitored volume decrease gives rise to the mass transfer coefficient $k_L$.
Figure~\ref{Fig:kL-all-conc} shows the resulting data for the different bubble diameters and the six different surfactant solutions plus the reference data obtained for the clean system.
Visual inspection shows a significant decrease of the mass transfer coefficient for bubbles rising in surfactant solution.
This mass transfer reduction is most prominent for the ellipsoidal bubbles in the diameter range up to $x_B=0.72$.
Evidently, for the bubble diameters in this range, the reduction factor is--within the bounds of the measurement errors--identical within the different groups comprising the
low, medium and high surfactant concentration, respectively. In other words, the experimental results immediately confirm the qualitative result that the mass transfer reduction factor is a function solely of the surface tension and independent of the type of surfactant.
The next section further substantiates the model performance by means of a quantitative description of the experimental data.
\section{Experimental validation of the novel mass transfer model}\label{sec4}
We are going to assess the model performance for the prediction of the mass transfer reduction, where we restrict our investigation
to the ellipsoidal bubbles, i.e.\ to the range of dimensionless bubble diameter up to $x_B=0.72$.
A fundamental problem to overcome is the fact that the experimental measurements only yield data on a global mass transfer coefficient, describing the bubble dissolution via an integral transfer rate that corresponds to an averaged local mass transfer rate.
To relate local mass transfer rates and their reduction to the bubble-averaged global values, some information on the surfactant distribution along the bubble surface is needed. For the ellipsoidal bubbles under consideration, the stagnant cap model provides this information.
The stagnant cap scenario also applies to non-spherical bubble as long as the bubble shape as well as the velocity field are axially symmetric and the process is quasi-stationary. Under these assumptions, the interfacial surfactant transport is governed by a substantially simplified
version of equation \eqref{E11}. In fact, the main contribution on the left-hand side is the convective transport. Hence, as $r^\Sigma =0$ for the considered case of an inert surfactant, the quasi-stationary surfactant profile approximately satisfies
\begin{equation}\label{eq:stagnant-cap1}
\na_\Sigma \cdot (c^\Sigma_i \vbf^\Sigma)=0.
\end{equation}
Passing to a co-moving frame, the velocity has vanishing normal (to the interface) component, as the shape is steady.
Furthermore, assuming zero swirl of the velocity field, the velocity at the bubble surface only has the meridional component that we denote as
$v_\theta^\Sigma$ in analogy to the spherical bubble case. Then \eqref{eq:stagnant-cap1} reduces to
\begin{equation}\label{eq:stagnant-cap2}
\frac{\partial}{\partial \theta} (c^\Sigma_i v_\theta^\Sigma)=0.
\end{equation}
Hence $c^\Sigma_i v_\theta^\Sigma = const$ and evaluation at one of the bubble poles, where $v_\theta^\Sigma = 0$, shows the constant to
equal zero. Consequently,
\begin{equation}\label{eq:stagnant-cap3}
c^\Sigma_i v_\theta^\Sigma=0.
\end{equation}
This in turn implies that, at every position, one of the factors vanishes. To sum up, we obtain
\begin{equation}\label{eq:stagnant-cap4}
c^\Sigma_i =0 \mbox{ for } 0\leq \theta \leq \theta_{cap} \quad\mbox{and}\quad
v_\theta^\Sigma =0 \mbox{ for } \theta_{cap} \leq \theta \leq \pi.
\end{equation}
Thus a certain fraction $f$ of the total bubble surface stays clean, while the remaining fraction $1-f$ is covered by surfactant.
While the surfactant concentration within the covered part is not homogeneous, it is known from analytical studies \cite{15} as well as from
numerical simulations \cite{takemura2005adsorption, kentheswaran2023impact} that the surfactant concentration shows a rather steep increase next to the cap angle before if approaches its maximum at the rear pole such that it does not vary strongly
on a huge portion of the covered area fraction.
Therefore, it is sensible to assume no mass transfer reduction on a certain fraction $f$ of the bubble surface
and a constant mass transfer reduction on the remaining part. Using the mass transfer reduction model according to \eqref{E65b}, we thus
aim to describe the experimental mass transfer data by the model
\begin{equation}\label{eq:bubble-mass-transfer-reduction1}
\frac{k_L^{\rm contam}}{k_L^{\rm clean}} = f + (1-f) e^{- \frac{\pi^\Sigma}{c^\Sigma_\infty RT}},
\end{equation}
where $\pi^\Sigma = \sigma^{\rm clean} - \sigma^{\rm contam}$ denotes the surface tension reduction (called 'surface pressure' in surface science) and $c^\Sigma_\infty$
characterises the maximum capacity of the surface to carry surfactant and can be viewed as a maximum possible surface coverage (in
moles per area). Note that the quantities $f$ and $c^\Sigma_\infty$ are unknown model parameters, which themselves will depend on the velocity field around the bubble, hence in particular on the rise velocity.

In a first step, we are going to fit\footnote{The least squares fit routine 'FindFit' in Wolfram Mathematica\textsuperscript{\textregistered} has been used.} the model parameters separately for bubbles of certain diameter (and, hence, rise velocity).
This will show whether the relation \eqref{eq:bubble-mass-transfer-reduction1} is, in principle, suitable to describe the measurements.
In a second step, we will then correlate the obtained model parameters to the rise velocity, aiming at a full model for the mass transfer
reduction of rising bubbles in the ellipsoidal regime.

\begin{figure}
\centering
\includegraphics[width=5.0in]{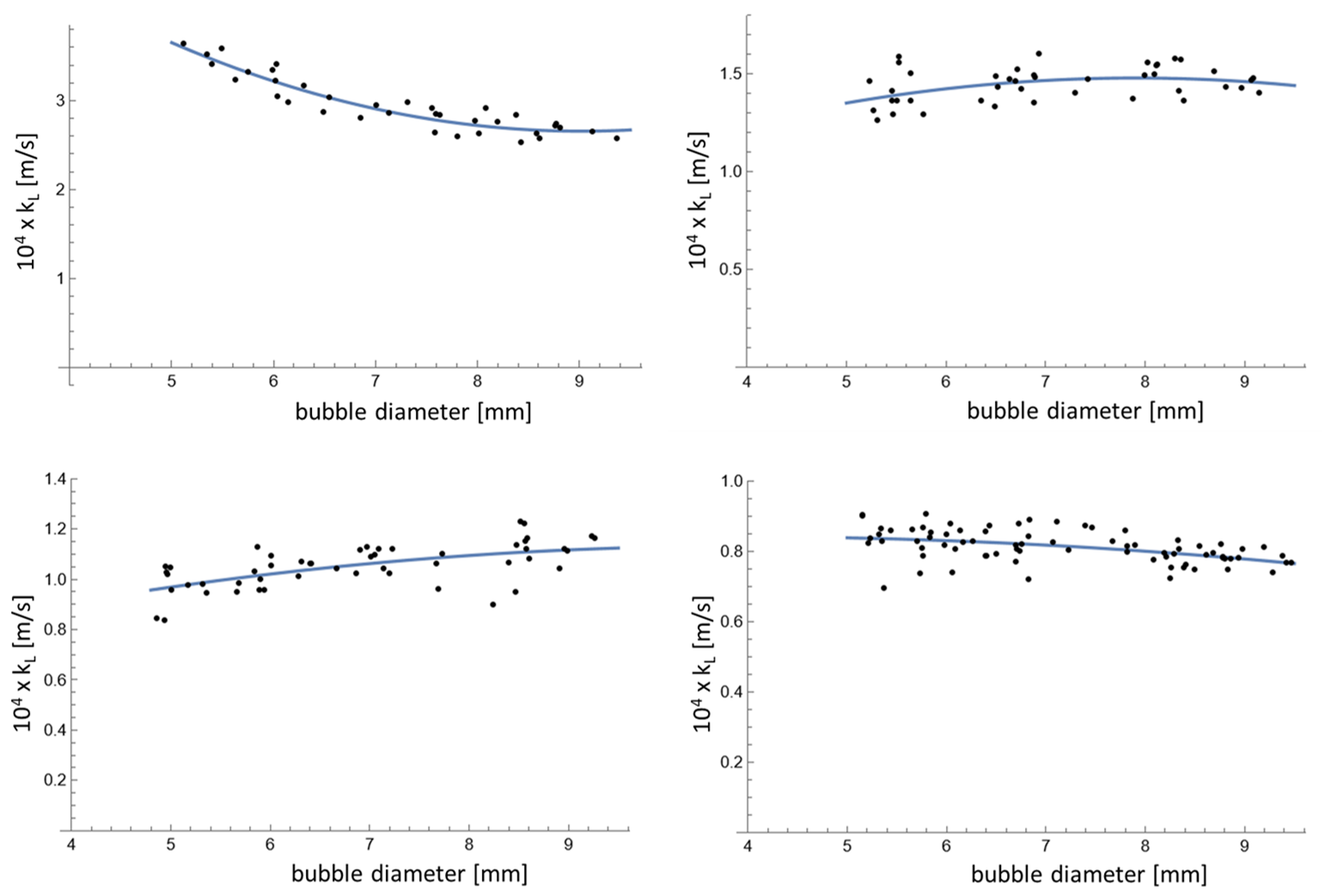}
\caption{Mass transfer coefficients from measurements and best quadratic fits for the clean system (top-left), low surfactant concentration (top-right), medium surfactant concentration (bottom-left) and high surfactant concentration (bottom-right).}\label{Fig:synthetic-kl}
\end{figure}

\begin{figure}
\centering
\includegraphics[width=5.0in]{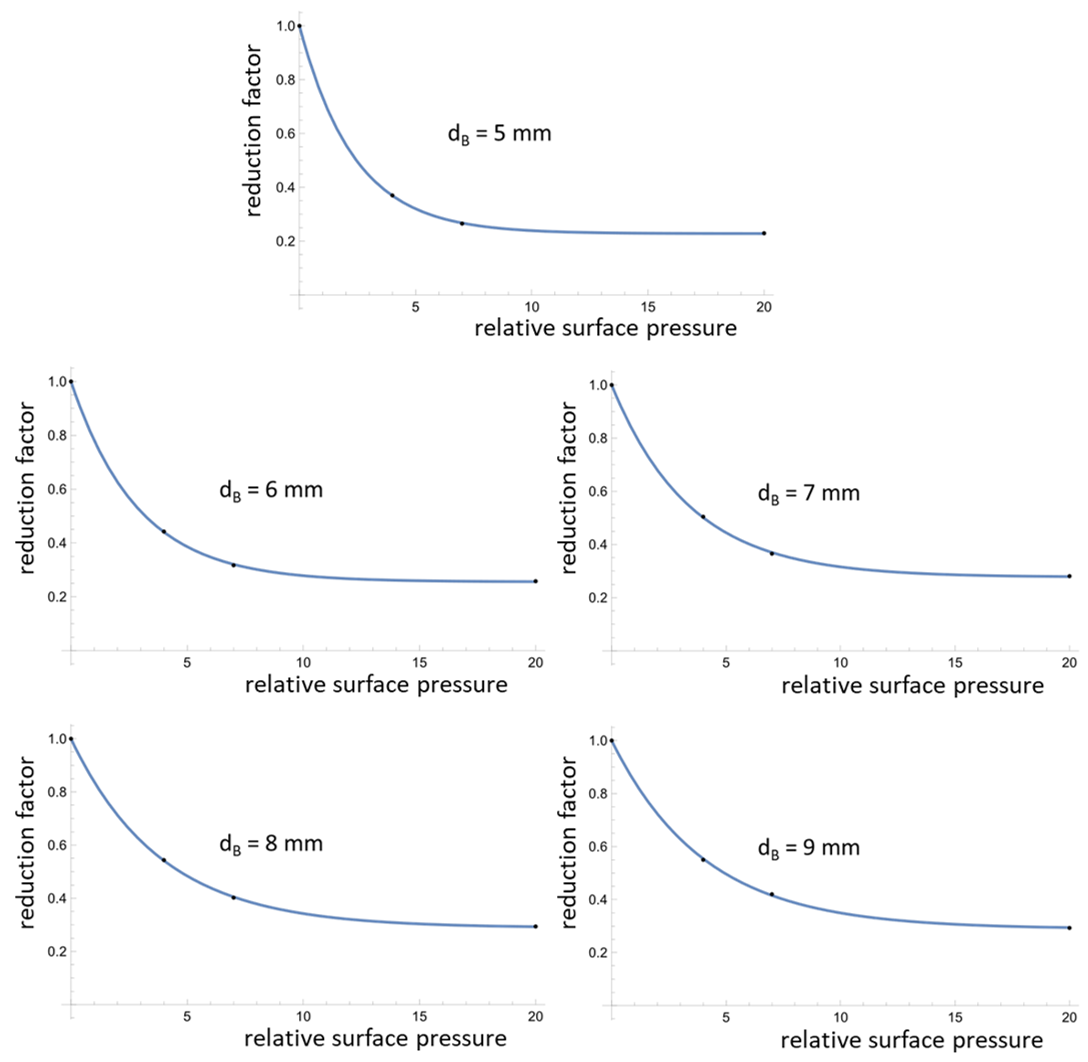}
\caption{Mass transfer reduction factor $k_L^{\rm contam}/k_L^{\rm clean}$ vs.\ relative surface pressure $\sigma^{\rm clean} - \sigma^{\rm contam}$
in mN/m. Comparison of fitted model with synthetic data.}\label{Fig:fit-results-individual}
\end{figure}

Another problem to overcome is that each individual experiment produces a certain bubble diameter that cannot be precisely preset but has to be measured. Hence there is no data available for precisely the same bubble diameter in the different surfactant solutions. To overcome this problem, we first need to derive such 'synthetic data' from the actual measurements. For this purpose, we collect the data of mass transfer coefficients for different bubble sizes within the groups of 'clean' (water) as well as 'low', 'medium' and 'high' surfactant concentrations.
These data sets are then fitted, where a simple quadratic fit turned out to describe the data satisfactory; see Fig.~\ref{Fig:synthetic-kl} for the quality of the fitting.
Using the quadratic fits, we generate synthetic data for the bubble diameters of $d_B=5$
, 6, 7, 8 and 9 mm.
\begin{table}[ht!]
\caption{Synthetic mass transfer reduction data obtained by quadratic fitting of the experimental results. The collected values
give the ratio $k_L^{\rm contam}/k_L^{\rm clean}$ for bubbles of the respective diameter and for the different surfactant solutions.}\label{tab:3}
\begin{tabular}{lllll}
\hline
diameter		&	clean			       	&	low 						&	medium 							&	high \\ \hline
5 mm			&	1.0				   		&	0.3699	\quad						&	0.2651 							&	0.2293\\ \hline
6 mm 			&	1.0				   		&	0.4424 							&	0.3170							&	0.2578\\ \hline
7 mm			&   1.0						&	0.5043 							&	0.3652 							&	0.2810\\ \hline
8 mm			&   1.0  					&	0.5436 							&	0.4022 							&	0.2939\\ \hline
9 mm			&   1.0						&	0.5500							&	0.4203 							&	0.2928\\ \hline
\end{tabular}
\end{table}
Tab.~\ref{tab:3} shows the resulting (synthetic) mass transfer reduction data.
Based on this data set, the model parameters $f$ and $c^\Sigma_\infty$ in the model \eqref{eq:bubble-mass-transfer-reduction1} have been fitted, where the data for each individual bubble diameter yields a corresponding pair $(f, c^\Sigma_\infty)$. The model describes the data very accurately as can be seen from Fig.~\ref{Fig:fit-results-individual}. 
The resulting model parameter values are listed in Tab.~\ref{tab:4}.
\begin{table}[ht!]
\caption{Best fit parameter values for the clean area fraction $f$ and the maximum concentration $c^\Sigma_\infty$ from fitting model
\eqref{eq:bubble-mass-transfer-reduction1} to the synthetic data.}\label{tab:4}
\begin{tabular}{lllll}
\hline
diameter		&	$f$	 [-]	       	   &	$c^\Sigma_\infty$ [$\mu$mol/m$^2$]	\\ \hline
5 mm			&	0.2280		\qquad		&	0.9646	\\ \hline
6 mm 			&	0.2555			   		&	1.1776  \\ \hline
7 mm			&   0.2771					&	1.3998 	\\ \hline
8 mm			&   0.2890 					&	1.5845 	\\ \hline
9 mm			&   0.2892					&	1.6615	\\ \hline
\end{tabular}
\end{table}
Both model parameters, $f$ and $c^\Sigma_\infty$, depend monotonically on the bubble diameter. 

\begin{figure}
\centering
\includegraphics[width=3.5in]{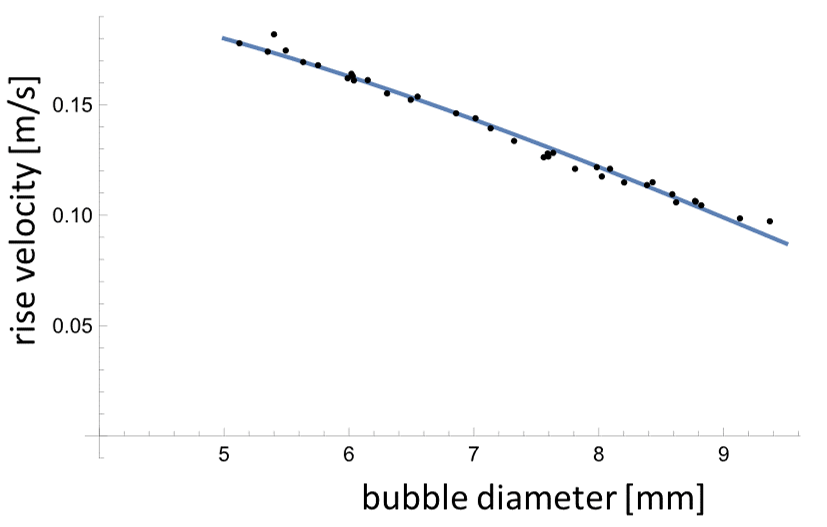}
\caption{Steady bubble rise velocity in water as a function of the bubble diameter in the ellipsoidal bubble range together with the model fit using equation \eqref{eq:vB-model1}.}\label{Fig:velo-fit-water}
\end{figure}

\begin{figure}
\centering
\includegraphics[width=5.0in]{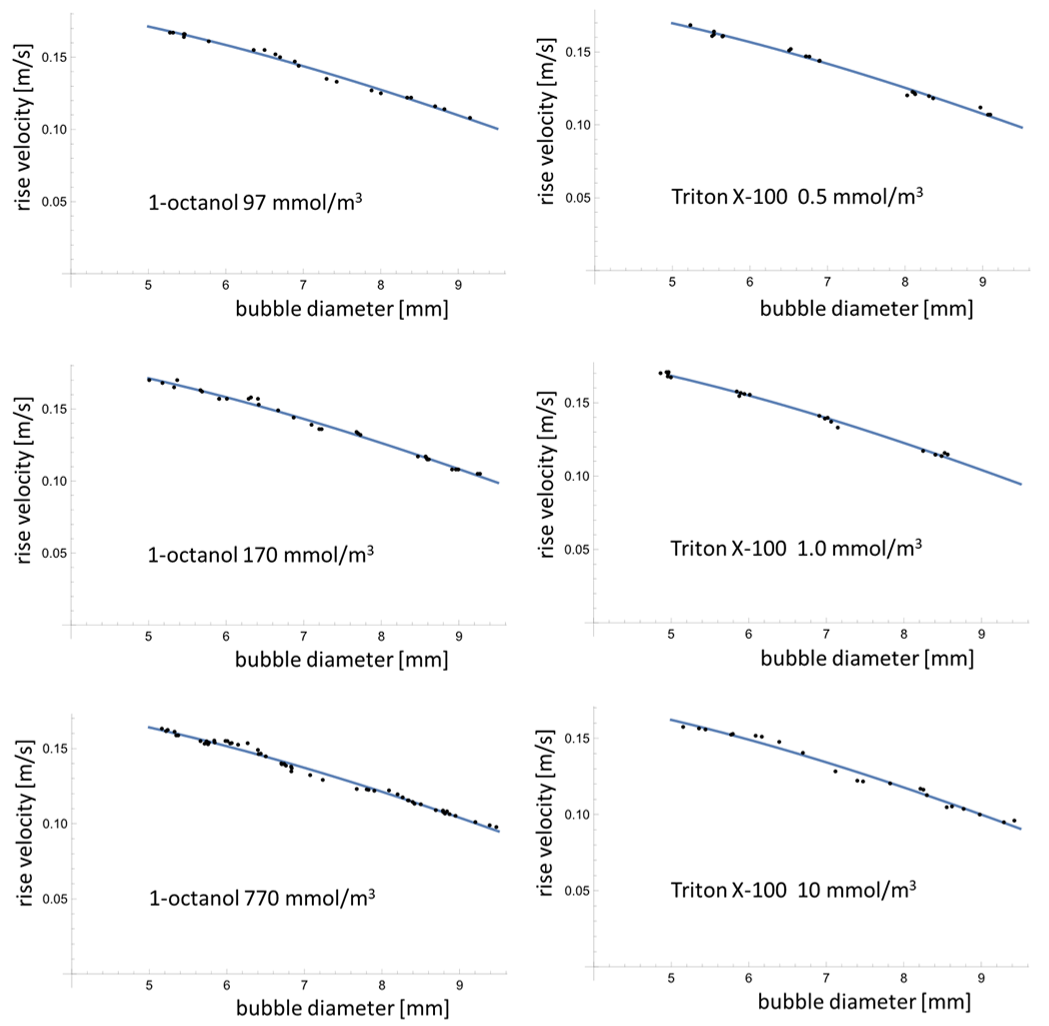}
\caption{Steady bubble rise velocity in water as a function of the bubble diameter in the ellipsoidal bubble range together with the model fit using equation \eqref{eq:vB-model1}.}\label{Fig:velo-fit-surfactant}
\end{figure}

\begin{figure}
\centering
\includegraphics[width=3.0in]{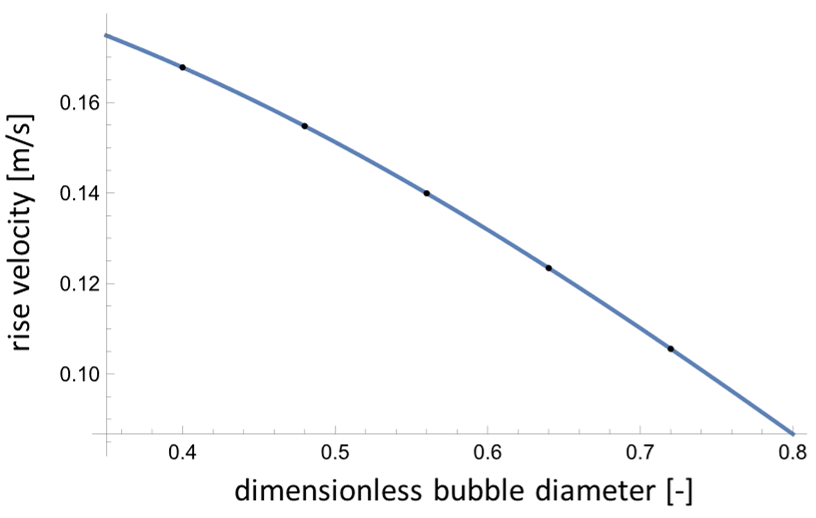}
\caption{Steady bubble rise velocity in any of the surfactant solutions as a function of the dimensionless bubble diameter in the ellipsoidal bubble range together with the model fit using equation \eqref{eq:vB-model1}.}\label{Fig:velo-fit-average}
\end{figure}

To obtain a complete model, applicable to rising bubbles of different diameter within the ellipsoidal range, we need to understand and model this parameter dependence. Evidently, the clean fraction $f$ of the bubble surface correlates with the angle of the stagnant cap, the so-called the cap angle. It is sensible that the cap angle is related to the stress at the bubble surface, as the stress pushes the adsorbed surfactant to the rear end of the bubble.
We estimate the stress to be roughly proportional to $\frac{v_B}{d_P-d_B}$, where $d_P$ denotes the pipe diameter.
This stress value will only act at a small belt around the bounding curve of the stagnant cap: Above this zone, the surface is clean and, hence, mobile, implying a much weaker velocity gradient. Below, the velocity field next to the bubble surface is about stagnant in the co-moving frame.
The extension of this 'belt' is unknown, but it should relate to the zone in which the surface tension changes from clean to contaminated value. From numerical simulations of several groups \cite{takemura2005adsorption, alke20093d, tukovic2012moving, pesci2018computational, kentheswaran2023impact} it is known that this transition indeed takes place within a rather small zone. 
Even if the precise extension of this zone is not known, the stress will hence act only on a part of the bubble surface,
the area of which is the length of the stagnant cap bounding curve times this unknown thickness.
Consequently, the total force that pushes the surfactant to the rear bubble surface part is proportional to the bubble diameter, rather than to the bubble surface.
From this consideration it follows that the clean fraction will, approximately, be proportional to 
\begin{equation}
\frac{d_B \, v_B}{d_P-d_B} = \frac{v_B}{1/x_B-1},
\end{equation}
where $x_B = d_B / d_P$ denotes the dimensionless bubble diameter.
This motivates to model the fraction $f$ of the clean surface by a relation of the (somewhat more general) type
\begin{equation}\label{eq_f-model1}
f = \frac{a\, v_B}{1/x_B-b}
\end{equation}
with model parameters $a,b >0$. 
We will use the equivalent relation
\begin{equation}\label{eq_f-model2}
\frac{v_B}{f} = \frac{\alpha}{x_B} - \beta 
\end{equation}
with model parameters $\alpha, \beta >0$. In order to fit the model parameters $\alpha, \beta$, we first need to also generate
synthetic velocity data, i.e.\ we first need to model the (steady) bubble rise velocity as a function of the bubble diameter.
For this purpose, we build on the know correlation $v_B \propto (1-x_B^2)^{3/2}$ from \cite{3b}, p.\ 233, and use the more general model
\begin{equation}\label{eq:vB-model1}
v_B = a \, \big( 1 - b \, x_B^2 \big)^{3/2}
\end{equation}
with two parameters $a,b >0$ instead of a single proportionality coefficient.
Even if we do not need to model the rise velocity for the clean system, we still show the result in Fig.~\ref{Fig:velo-fit-water} to see the quality of the fit. The fitted parameter values are $a = 0.2211$ and $b = 0.8005$.
%
\begin{table}[ht!]
\caption{Synthetic mass transfer reduction data obtained by quadratic fitting of the experimental results. The collected values
give the ratio $k_L^{\rm contam}/k_L^{\rm clean}$ for bubbles of the respective diameter and for the different surfactant solutions.}\label{tab:5}
\begin{tabular}{lllll}
\hline
$c_{\rm surf}$ [mmol/m$^3$] 		&	$a$	 [-]	       	   &	$b$	 [-]		\\ \hline
97 (1-octanol)		    	&	0.2016		\qquad		&	0.6431	\\ \hline
170 (1-octanol) 			&	0.2024			   		&	0.6582  \\ \hline
770 (1-octanol)		       	&   0.1936					&	0.6552 	\\ \hline
0.5 (Triton X-100)			&   0.2003 					&	0.6551 	\\ \hline
1.0 (Triton X-100)			&   0.2000 					&	0.6810 	\\ \hline
10 (Triton X-100)			&   0.1929 					&	0.6848 	\\ \hline
\end{tabular}
\end{table}
The rise velocities in the surfactant solutions can be approximated by the same model with very good accuracy. Tab.~\ref{tab:5}
lists the parameter values obtained from the fitting and Fig.~\ref{Fig:velo-fit-surfactant} depicts the model outcome and the underlying data.
As the rise velocity correlations for the surfactant solutions are all very close to each other, we can use a common velocity-to-diameter relation. To obtain the latter, we use the above individual fittings from which we compute synthetic rise velocities for the used set
of synthetic bubble diameters of  $d_B=5$, 6, 7, 8 and 9 mm.
The resulting rise velocities, very similar for identical bubble diameter, are then averaged.
This yields the synthetic data for the rise velocities in surfactant solution as shown in Tab.~\ref{tab:6}.
\begin{table}[ht!]
\caption{Synthetic averaged rise velocities for bubbles in any of the surfactant solution.}\label{tab:6}
\begin{tabular}{lllll}
\hline
diameter		&	$v_B$	 [m/s]	  \\ \hline
5 mm			&	0.1678		\\ \hline
6 mm 			&	0.1548		 \\ \hline
7 mm			&   0.1399			\\ \hline
8 mm			&   0.1234 			\\ \hline
9 mm			&   0.1056			\\ \hline
\end{tabular}
\end{table}
To be able to include these data into the complete model, we again fit the synthetic data using the model \eqref{eq:vB-model1}.
The best fit is obtained for parameter values $a = 0.1985$ and $b = 0.6627$. Fig.~\ref{Fig:velo-fit-average} displays the result. This model for the rise velocity as a function of the bubble diameter, with fixed given
parameter values, will be a building block inside the complete model.

\begin{figure}
\centering
\includegraphics[width=3.0in]{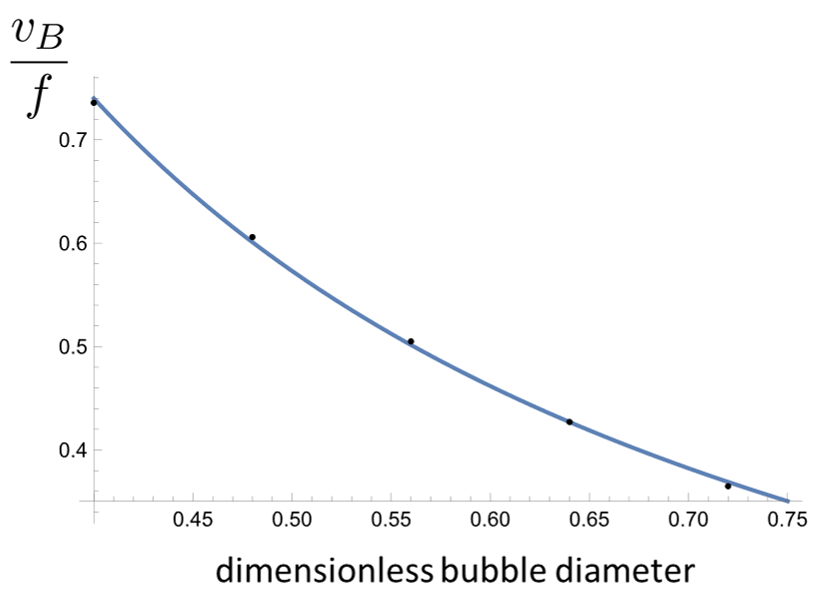}
\caption{$v_B/f$ in m/s in dependence of the dimensionless bubble diameter $x_B$ together with the model fit using equation \eqref{eq_f-model2}.}\label{Fig:vBf-fit}
\end{figure}

Having the rise velocities from Tab.~\ref{tab:6}, we can now resume correlating $v_B/f$ to the dimensionless bubble diameter $x_B$ via equation \eqref{eq_f-model2}. The model fit yields $\alpha = 0.3340$ and $\beta = 0.0948$ m/s for the parameters, and the comparison between model and the data is displayed in Fig.~\ref{Fig:vBf-fit}.

\begin{figure}
\centering
\includegraphics[width=3.0in]{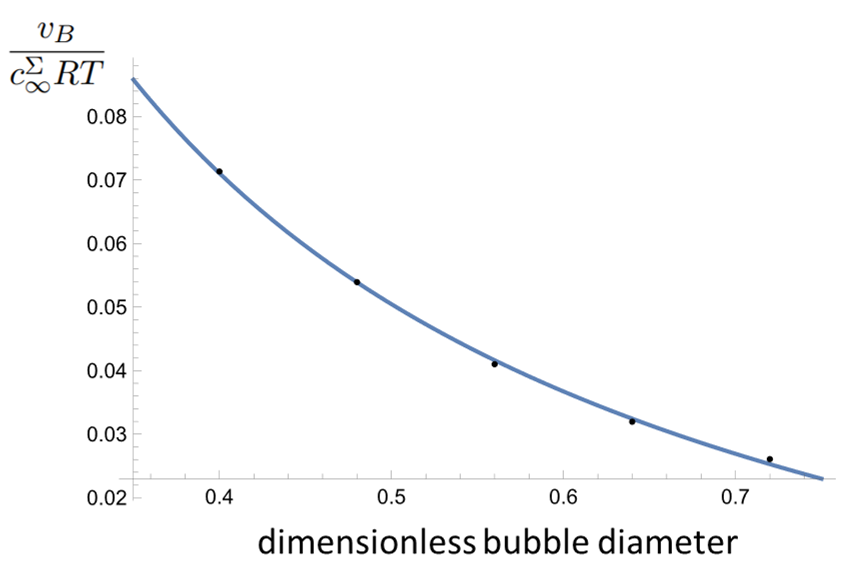}
\caption{$v_B/c_\infty^\Sigma RT$ in mm s/kg in dependence of the dimensionless bubble diameter $x_B$ together with the model fit using equation \eqref{eq:cmax-model}.}\label{Fig:cmax-fit}
\end{figure}

A very similar reasoning applies to the maximum surface capacity $c_\infty^\Sigma$, respectively the corresponding surface pressure
$c_\infty^\Sigma RT$ of the maximal compressed surface coverage under the respective stress conditions. 
Indeed, the model
\begin{equation}\label{eq:cmax-model}
\frac{v_B}{c_\infty^\Sigma RT} = \frac{\gamma}{x_B} - \delta 
\end{equation}
can be accurately fitted to the data as shown by the result displayed in Fig.~\ref{Fig:cmax-fit}.
The fitted parameter values are $\gamma = 0.0412$ and $\delta = 0.0320$, both with unit mm s/kg.

\begin{figure}
\centering
\includegraphics[width=5.0in]{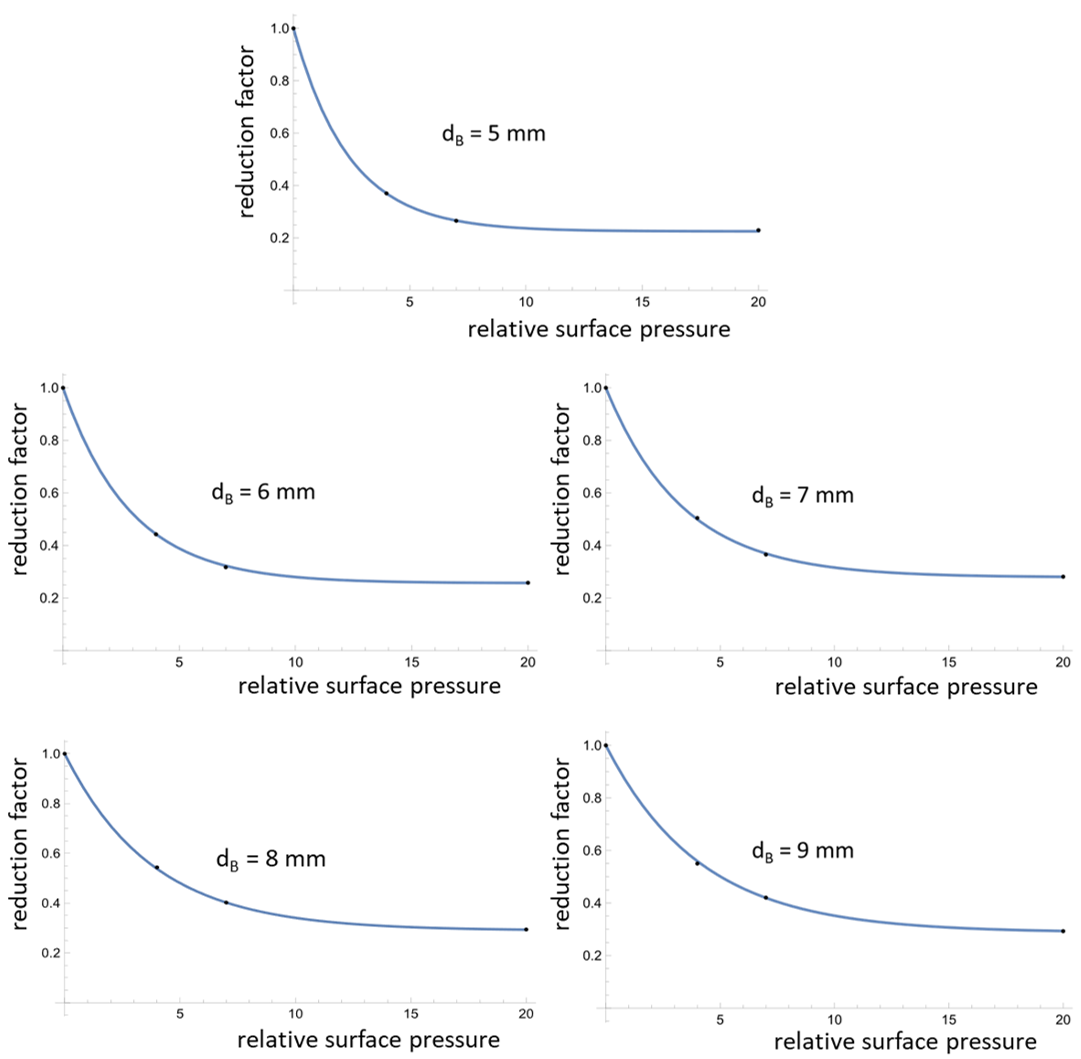}
\caption{Mass transfer reduction factor $k_L^{\rm contam}/k_L^{\rm clean}$ vs.\ relative surface pressure $\sigma^{\rm clean} - \sigma^{\rm contam}$
in mN/m. Comparison of fitted overall model with synthetic data.}\label{Fig:fit-results-joint}
\end{figure}

Having all the parameter dependencies accurately modeled, these building blocks are now inserted into the mass transfer reduction model
\eqref{eq:bubble-mass-transfer-reduction1}. As mentioned above, we keep the velocity-to-diameter dependence fixed, where the leading coefficient is absorbed into the other parameters. This results in the overall model
\begin{equation}\label{eq:bubble-mass-transfer-reduction-full}
\frac{k_L^{\rm contam}}{k_L^{\rm clean}} = \frac{( 1 - \lambda \, x_B^2 )^{3/2}}{a+\frac{b}{x_B}}
+ \Big( 1 - \frac{( 1 - \lambda \, x_B^2 )^{3/2}}{a+\frac{b}{x_B}} \Big) 
e^{- \frac{(c + d/x_B)\, \pi^\Sigma /\sigma_0}{( 1 - \lambda \, x_B^2 )^{3/2}}},
\end{equation}
where the (relative) surface pressure is made dimensionless with the surface tension of pure water, $\sigma_0 = 72.5$ mN/m, and $\lambda = 0.6627$ is fixed from the prior rise velocity fitting. We then use the separately found parameter values from above as initial values for a final parameter fit, in which the four dimensionless model parameters $a,b,c,d$ are fitted to the complete synthetic data set.
The fitted values are $a=-0.5420$, $b=1.7207$, $c=-11.0$, and $d=14.78$.
The model describes the data as accurate as with the individual fitting from above. For comparison, the results
are shown in Fig.~\ref{Fig:fit-results-joint}, where one can hardly see a difference between Fig.~\ref{Fig:fit-results-individual} and Fig.~\ref{Fig:fit-results-joint}.

\section{Conclusion and Outlook}
The successful experimental validation of the novel mass transfer model proves that local mass transfer hindrance due to the presence of surfactant can be accurately described by continuum thermodynamics within the sharp-interface framework. For this it is essential that area-specific surface concentrations of all constituents, i.e.\ surfactant species as well as transfer species, are included in the model. This is a prerequisite in order to capture the full mixture thermodynamics on the interface, where the surface pressure mediates a strong coupling via the dependence of surface chemical potentials on surface pressure. For this purpose, it is not the inertia of the molecules in the interface layer but their molar mass, which needs to be accounted for.

The resulting model resembles Langmuir's energy barrier model, but now in a local form and as part of a thermodynamically consistent overall model.
This allows to combine the local mass transfer reduction model with knowledge on heterogeneous surface coverage by surfactant to quantitatively describe the integral mass transfer across the full bubble surface. This bubble-mass transfer model perfectly describes the experimental measurements after appropriate parameter values are chosen by standard parameter fitting.

The model fit in the present investigation has to use synthetic data, as bubbles of pre-specified diameter cannot be generated experimentally.
It would be interesting to employ a Bayesian modeling approach that could work with the experimental data as is. A corresponding workflow has been developed in \cite{gossel2025scale} in a different context and shall be applied to mass transfer reduction in future work.

An interesting next step would be a possible extension of the overall model to cover the Taylor bubble regime, too.
Furthermore, an implementation of the local mass transfer reduction model into numerical methods for resolved mass transfer simulations would allow 
for deep insights into the interplay between local hydrodynamics, influenced by surfactant-induced Marangoni effects, 
surfactant coverage and local/global mass transfer rates.
A far reaching step would be the extension of the model to cover ionic surfactant, salt effects and highly concentrated bulk mixtures etc.,
as charge effects of ionic species need to be accounted for, including electromigrative transport contributions.

Finally, for modeling mass transfer in large scale contact apparatuses under the influence of contaminations or surface active additives, simplified models are demanded such as two-film models accounting for hindered mass transfer.
\section*{Acknowledgment}
The first author gratefully acknowledges financial support provided by the German Research Foundation (DFG) within the Priority Program SPP1740 ``Reactive Bubbly Flows'' (project ID 256739956).

%
%
%
%
%
%
%
\end{document}